\documentclass[aps,prl,twocolumn,showpacs,superscriptaddress]{revtex4-1}

\usepackage{graphicx}
\usepackage[colorlinks,citecolor=blue,urlcolor=blue,linkcolor=blue]{hyperref}
\usepackage{siunitx}	
\usepackage{amsmath}
\usepackage{amssymb}
\usepackage{amsthm}
\usepackage{braket}
\usepackage{booktabs}
\usepackage{footmisc}
\usepackage{mathrsfs}

\newcommand{\unit}{1\!\!1}

\newcommand{\modu}{\mathrm{mod}}
\newcommand{\perm}{\mathrm{perm}}
\newcommand{\abs}[1]{\left| #1 \right|} 
\newcommand{\bracket}[2]{\ensuremath{\left\langle#1 \vphantom{#2}\right| \left. #2\vphantom{#1}\right\rangle}}

\begin{document}
\title{Totally Destructive Many-Particle Interference}

\author{Christoph Dittel}
\email{christoph.dittel@uibk.ac.at}
\affiliation{Institut f{\"u}r Experimentalphysik, Universit{\"a}t Innsbruck, Technikerstr. 25, 6020 Innsbruck, Austria}
\affiliation{Physikalisches Institut, Albert-Ludwigs-Universit{\"a}t Freiburg, Hermann-Herder-Str. 3, 79104 Freiburg, Germany}

\author{Gabriel Dufour}
\affiliation{Physikalisches Institut, Albert-Ludwigs-Universit{\"a}t Freiburg, Hermann-Herder-Str. 3, 79104 Freiburg, Germany}
\affiliation{Freiburg Institute for Advanced Studies, Albert-Ludwigs-Universit{\"a}t-Freiburg, Albertstr. 19, 79104 Freiburg, Germany}

\author{Mattia Walschaers}
\affiliation{Laboratoire Kastler Brossel, UPMC-Sorbonne Universit\'{e}s, ENS-PSL Research University, Coll\`{e}ge de France, CNRS; 4 place Jussieu, 75252 Paris, France}

\author{Gregor Weihs}
\affiliation{Institut f{\"u}r Experimentalphysik, Universit{\"a}t Innsbruck, Technikerstr. 25, 6020 Innsbruck, Austria}

\author{Andreas Buchleitner}
\affiliation{Physikalisches Institut, Albert-Ludwigs-Universit{\"a}t Freiburg, Hermann-Herder-Str. 3, 79104 Freiburg, Germany}

\author{Robert Keil}
\affiliation{Institut f{\"u}r Experimentalphysik, Universit{\"a}t Innsbruck, Technikerstr. 25, 6020 Innsbruck, Austria}


\date{\today}

\begin{abstract}
In a general, multi-mode scattering setup, we show how the permutation symmetry of a many-particle input state determines those scattering unitaries which exhibit strictly suppressed many-particle transition events. We formulate purely algebraic suppression laws that identify these events and show that the many-particle interference at their origin is robust under weak disorder and imperfect indistinguishability of the interfering particles. Finally, we demonstrate that all suppression laws so far described in the literature are embedded in the general framework that we here introduce.
\end{abstract}


\maketitle


Many-particle interference is a distinguished feature of quantum theory. It is grounded in the coherent evolution of many-particle states and provides a resource for applications ranging from quantum metrology \cite{Walther-DG-2004} to quantum computation \cite{Knill-SE-2001,OBrien-OQ-2007} and quantum information \cite{Shih-NT-1988,Halder-EI-2007}. Its first experimental verification dates back to 1987, when Hong, Ou and Mandel (HOM) \cite{Hong-MS-1987} observed the quantum statistical signature of two interfering bosons by injecting one photon into each input port of a balanced beam splitter. The bosonic nature of the photons \cite{Shih-NT-1988,Loudon-FB-1998} causes totally destructive interference of the two-particle amplitudes leading to one particle in each output port, so that both particles leave the beam splitter through the same port. The same phenomenon can also be observed in other bosonic systems, such as trapped $^{87}\mathrm{Rb}$ atoms in a double-well potential \cite{Kaufman-TW-2014} or overlapping beams of metastable $^4\mathrm{He}$ atoms \cite{Lopes-AH-2015}. For indistinguishable fermions, on the other hand, Pauli's exclusion principle \cite{Pauli-EP-1964} forbids the simultaneous occupation of the same physical state. Consequently, two fermions never leave the beam splitter through the same port, as demonstrated in electron collisions \cite{Liu-QI-1998,Kiesel-OH-2002}.

An important aspect of HOM interference is the fact that the beam splitter is balanced, that is, it features equal single-particle probabilities for transmission and reflection as well as a particular phase relation arising from the conservation of energy \cite{Zeilinger-GP-1981}. Consequently, the probability amplitudes for both photons being transmitted and both photons being reflected cancel each other perfectly. We thus naturally ask the question whether and how totally destructive interference may arise in more complex scenarios with larger numbers of particles and input/output ports (henceforth termed modes). Efforts towards solving this question were undertaken by feeding two particles into beam splitters with three \cite{Mattle-NC-1995,Weihs-TP-1996,Meany-NC-2012} and four \cite{Mattle-NC-1995,Meany-NC-2012} modes, via theoretical \cite{Campos-TP-2000,Guise-CL-2014} and experimental \cite{Spagnolo-TP-2013,Tillmann-GM-2015,Menssen-DM-2017} extensions to three particles entering a tritter, as well as experiments demonstrating bosonic bunching for four \cite{Ou-OF-1999,Xiang-DT-2006} and six \cite{Xiang-DT-2006,Niu-OG-2009} photons and two modes. In all these scenarios, destructive interference was observed for certain input-output configurations. Regarding a generalisation of the HOM effect to arbitrary particle or mode numbers, destructive interference was studied in free-space propagation \cite{Maehrlein-HOM-2017}, and several individual unitary transformations were investigated, for which a large number of final many-particle output events are suppressed due to totally destructive interference: the discrete Fourier transformation \cite{Lim-GH-2005,Tichy-ZT-2010,Tichy-MP-2012,Carolan-UL-2015,Crespi-SL-2016,Su-MI-2017}, the $J_x$ unitary \cite{Perez-Leija-CQ-2013,Weimann-IQ-2016}, Sylvester interferometers \cite{Crespi-SL-2015,Viggianiello-EG-2017} and hypercube unitaries \cite{Dittel-MB-2017}. In all these cases, which can be interpreted as different generalisations of the two-port beam splitter, so called \textit{suppression laws} have been formulated. However, to which extent all these distinct interference scenarios can be understood as the consequence of one underlying principle, has remained an unsolved question to date.

Here we uncover this very principle. While all previous scenarios \cite{Lim-GH-2005,Tichy-ZT-2010,Tichy-MP-2012,Perez-Leija-CQ-2013,Weimann-IQ-2016,Crespi-SL-2015,Viggianiello-EG-2017,Dittel-MB-2017} considered \emph{specific} transformation matrices to infer the associated set of output events fully suppressed by destructive many-particle interference, we here derive entire \emph{classes} of transformation matrices, together with the associated suppression laws, solely from the permutation symmetries of arbitrary many-particle input states. We show how to construct these unitaries and relate the prediction of suppressed many-particle transmission events to the simple evaluation of an associated set of eigenvalues. This formalism unifies all known cases of many-particle suppression laws under one general perspective and allows targeted design of multi-mode multi-particle setups. For some of the already established suppression laws, our general result even enlarges the set of suppressed events. We provide a generic example by application of our formalism to the discrete Fourier transformation, for which our result coincides with the findings in \cite{Tichy-MP-2012} for bosons, and extends them for fermions. Furthermore, we show that totally destructive interference is robust with respect to partial particle distinguishability and small deviations from the ideal unitary. For an extensive discussion on technical aspects, various applications, and extensions of these principles we refer the reader to \cite{Dittel-TDarticle-2017}.


We consider the transmission of a Fock-state of $N$ non-interacting particles across an $n$-mode scattering device. The configuration of a many-particle state with $r_j$ particles injected in mode $j\in\{1,\dots,n\}$ is denoted by the initial \textit{mode occupation list} $\vec{r}=(r_1,\dots,r_n)$ and its corresponding \textit{mode assignment list} $\vec{d}(\vec{r})=(d_1(\vec{r}),\dots,d_N(\vec{r}))$, with $d_\alpha(\vec{r})$ specifying the mode occupied by the $\alpha$th particle, $\alpha\in\{1,\dots,N\}$ \cite{Tichy-Thesis-2011,Tichy-MP-2012,Tichy-II-2014,Dittel-MB-2017}. Analogously, the final particle configuration is denoted by $\vec{s}$ and $\vec{d}(\vec{s})$. Unless otherwise stated, we consider all particles as mutually indistinguishable bosons (B) or fermions (F) such that the ordering in $\vec{d}(\vec{r})$ and $\vec{d}(\vec{s})$ is irrelevant.

Given the single-particle unitary $U\in\mathbb{C}^{n\times n}$, with $U_{j,k}$ the probability amplitude for a transition from initial mode $j$ to final mode $k$, any many-particle transformation is governed by the scattering matrix $M\in\mathbb{C}^{N\times N}$, with $M_{\alpha,\beta}\equiv U_{d_\alpha(\vec{r}),d_\beta(\vec{s})}$ composed of those rows and columns of $U$ that correspond to occupied initial and final modes, respectively. The transition probability for bosons is related to the permanent of the scattering matrix, $P_{\mathrm{B}}( \vec{r},\vec{s},U)\propto \abs{\perm\left(M\right)}^2$, whereas for fermions it is given by the determinant, $P_{\mathrm{F}}( \vec{r},\vec{s},U)= \abs{\det\left(M\right)}^2$ \cite{Mayer-CS-2011,Tichy-Thesis-2011,Tichy-MP-2012,Tichy-II-2014,Crespi-SL-2015,Dittel-MB-2017}. For distinguishable particles (D), on the other hand, the transition probability is insensitive to scattering phases and readily obtained from the single particle transition probabilities $|U_{j,k}|^2$. It obeys $P_{\mathrm{D}}( \vec{r},\vec{s},U)\propto \perm(|M|^2)$ \cite{Mayer-CS-2011,Tichy-Thesis-2011,Tichy-MP-2012,Tichy-II-2014,Crespi-SL-2015,Dittel-MB-2017} where $|.|^2$ denotes the element-wise modulus squared.

A suppression of final particle configurations due to totally destructive interference must naturally result from phase relations in $M$ that lead to a cancellation of all terms in the determinant or permanent. Given an initial particle configuration $\vec{r}$, we therefore investigate the characteristics of transformation matrices $U$ that exhibit such phase relations. Because the transition probability involves contributions from all permutations of the initial (or final) particle configuration, intuitively, these unitaries are closely connected to the permutation characteristics of $\vec{r}$ (or $\vec{s}$). 

Consequently, let $\pi\in \mathrm{S}_n$, with $\mathrm{S}_n$ the symmetric group on the set of modes $\{1,\dots,n\}$, denote a non-trivial permutation of all input modes that leaves the initial particle configuration unchanged. Accordingly, the mode occupation list $\vec{r}$ is invariant under the corresponding permutation operator $\mathscr{P}$,
\begin{align}\label{eq:invariant}
\mathscr{P}\vec{r}=\vec{r},
\end{align}
with $\mathscr{P}_{j,k}=\delta_{\pi(j),k}$. An eigendecomposition of the permutation operator reads $\mathscr{P}=ADA^\dagger$, where the columns of the unitary matrix $A$ correspond to the eigenvectors of $\mathscr{P}$, and $D=\mathrm{diag}(\lambda_1,\dots,\lambda_n)$ lists the eigenvalues $\lambda_j$ associated with the $j$th eigenvector. As $A$ can be any eigenbasis of $\mathscr{P}$, its columns can be permuted arbitrarily (accompanied by the corresponding reordering of $D$) and for each $q$-fold degenerate eigenvalue one can freely choose a basis of the corresponding $q$-dimensional eigenspace.

If the order of $\pi$ is $m$, i.e., $\mathscr{P}^m=\unit$, the eigenvalues listed in $D$ are $m$th roots of unity. More precisely, the eigenvalues are determined by the cycle lengths of $\pi$: for each cycle with length $l$, all $l$th roots of unity appear in $D$. For example, the permutation $\pi=(1~2~3)(4~5~6)(7~8)$ consists of two cycles of length $3$ and one cycle of length $2$, and the corresponding permutation operator $\mathscr{P}$ has eigenvalues $\{e^{i\frac{2\pi}{3}},e^{i\frac{4\pi}{3}},1,e^{i\frac{2\pi}{3}},e^{i\frac{4\pi}{3}},1,-1,1\}$.

As we show in the following, for initial states satisfying Eq.~\eqref{eq:invariant} and any choice of eigenbasis $A$, any unitary matrix
\begin{align}\label{eq:Ueigen}
U=\Theta~A~\Sigma
\end{align}
exhibits a large number of suppressed final particle configurations. Note that $U$ is specified by the permutation characteristics of the initial state, and $\Theta\in\mathbb{C}^{n\times n}$ and $\Sigma\in\mathbb{C}^{n\times n}$ are arbitrary unitary diagonal matrices accounting for local phases of initial and final modes, respectively,  which do not affect many-particle interference. 

To infer the suppressed output events for unitaries of the form~\eqref{eq:Ueigen} we utilize the latters' relation with the permutation operator $\mathscr{P}$. It can be straightforwardly verified that the action of $\mathscr{P}$ leaves these unitaries unaffected except for local phases:
\begin{align}\label{eq:MatrixSym}
\mathscr{P}~U=Z~U~D,
\end{align}
with $Z=\mathscr{P}~\Theta~\mathscr{P}^\dagger~\Theta^\dagger$. Note that $D$ is diagonal and unitary by definition, and $Z$ inherits these properties from $\Theta$ through the permutations induced by $\mathscr{P}$. Equation~\eqref{eq:MatrixSym} reveals that the permuted rows of $U$ (which enter the permanent/determinant) only differ by local phases, defined by $Z$ and $D$. Since a vanishing transition probability must be independent of $Z$ (which originates from arbitrary local phases in $\Theta$), Eq.~\eqref{eq:MatrixSym} also reveals that totally destructive interference can only depend on the eigenvalues in $D$.

\begin{figure}[t]
\centering
\includegraphics[width=0.49\textwidth]{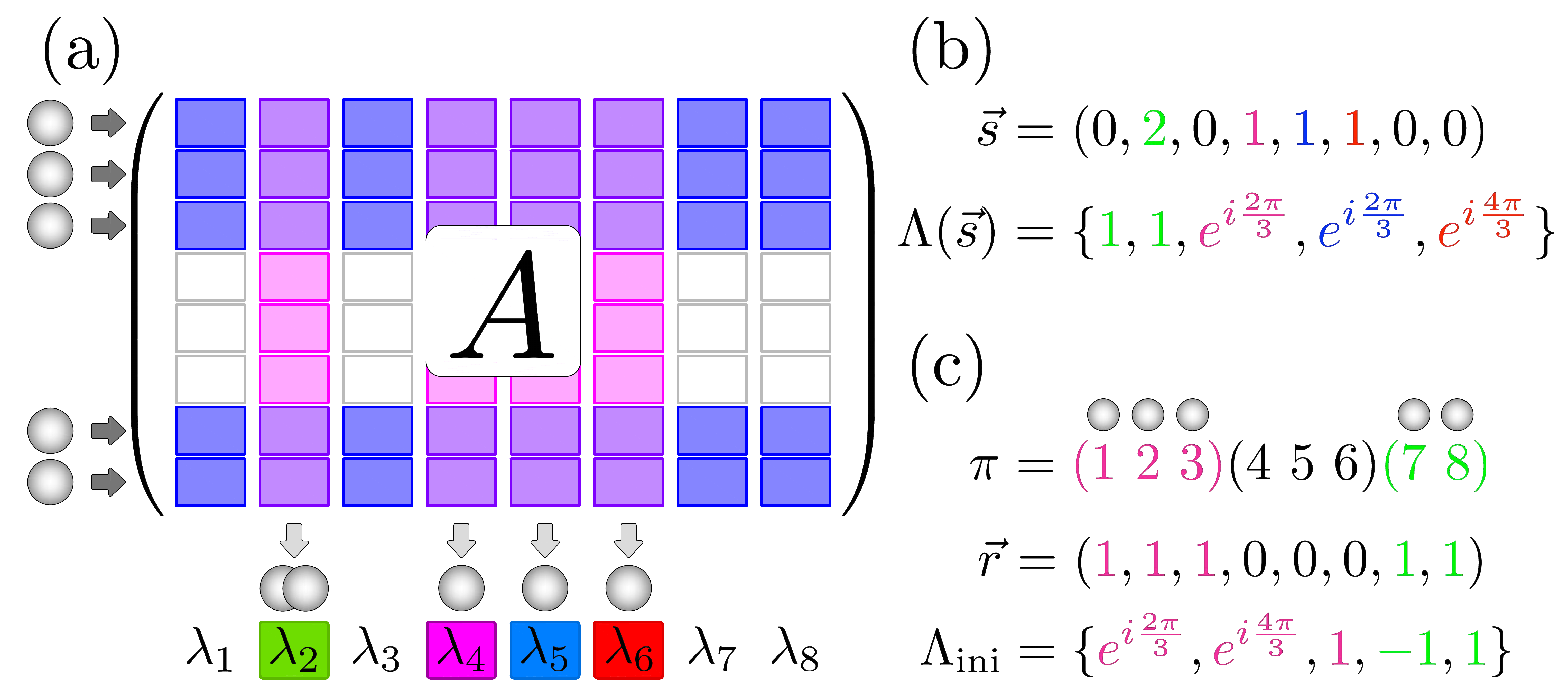} 
\caption{(Color online) Initial and final eigenvalue distribution for $N=5$ particles injected into $n=8$ modes, with input and output mode occupation lists $\vec{r}$ and $\vec{s}$, respectively. The input $\vec{r}$ is invariant under the permutation $\pi=(1~2~3)(4~5~6)(7~8)$, with eigenvalue matrix $D=\mathrm{diag}(\lambda_1,\dots,\lambda_n)$, $\lambda_1=\lambda_2=\lambda_3=1$, $\lambda_4=\lambda_5=e^{i\frac{2\pi}{3}}$, $\lambda_6=\lambda_7=e^{i\frac{4\pi}{3}}$ and $\lambda_8=-1$. The mode assignment list on output, $\vec{d}(\vec{s})=(2,2,4,5,6)$, then defines the associated final eigenvalue distribution by the corresponding mapping of $\{\lambda_1,\dots,\lambda_n\}$ onto $\Lambda(\vec{s})=\{\lambda_{d_1(\vec{s})},\dots,\lambda_{d_N(\vec{s})}\}=\{1,1,e^{i\frac{2\pi}{3}},e^{i\frac{2\pi}{3}},e^{i\frac{4\pi}{3}}\}$. The initial eigenvalue distribution $\Lambda_\mathrm{ini}$, defined by $\vec{r}$ together with the cycle lengths of $\pi$, is a subset of $\{\lambda_1,\dots,\lambda_n\}$: $\Lambda_\mathrm{ini}=\{e^{i\frac{2\pi}{3}},e^{i\frac{4\pi}{3}},1,-1,1\}$ (see main text for the definition of $\Lambda_\mathrm{ini}$).}
\label{fig:eigenvalues}
\end{figure}

As we show below, forbidden output configurations can be determined by the  \textit{final eigenvalue distribution} $\Lambda(\vec{s})=\{\lambda_{d_1(\vec{s})},\dots,\lambda_{d_N(\vec{s})}\}$, which contains all $N$ eigenvalues $\Lambda_\alpha(\vec{s})=\lambda_{d_\alpha(\vec{s})}$ associated with the final mode assignment list $\vec{d}(\vec{s})$, as illustrated in Figs.~\ref{fig:eigenvalues}~(a) and~(b). Note that $\Lambda(\vec{s})$ is a multiset, that is, multiple instances of its elements are allowed and the order is irrelevant. For indistinguishable fermions, we further introduce the \textit{initial eigenvalue distribution} $\Lambda_\mathrm{ini}$, which is predetermined by the initial mode occupation $\vec{r}$ and contains a subset of those eigenvalues listed in $D$: In consideration of Pauli's principle, Eq.~\eqref{eq:invariant} holds if and only if all modes belonging to the same cycle of $\pi$ are initially empty or populated by exactly one fermion. Then, each cycle with length $l$ whose modes are initially populated contributes the set of eigenvalues $e^{i2\pi\frac{k}{l}}$ with $k=1,\dots,l$ to the multiset $\Lambda_\mathrm{ini}$ (see Fig.~\ref{fig:eigenvalues}~(c) for the previously discussed example). Whether $\vec{s}$ is suppressed can then be determined by the following suppression laws, which we find after involved technical manipulations, which are detailed in \cite{Dittel-TDarticle-2017}. \\

\textbf{Suppression Laws:}
Let $\vec{r}$ be any initial particle configuration of $N$ indistinguishable particles which is invariant under the mode-permutation operation $\mathscr{P}$ defined by a permutation $\pi\in \mathrm{S}_n$. Let $ADA^\dagger$ be any eigendecomposition of $\mathscr{P}$ with eigenvalues $D=\mathrm{diag}(\lambda_1,\dots,\lambda_n)$ and eigenbasis $A\in\mathbb{C}^{n\times n}$, $\Lambda_\mathrm{ini}$ and $\Lambda(\vec{s})$ the initial and final eigenvalue distribution, respectively, and $\Theta$ and $\Sigma$ arbitrary diagonal unitary matrices. For the unitary transformation matrix $U=\Theta~A~\Sigma$, final particle configurations $\vec{s}$ are suppressed, i.e. $P(\vec{r},\vec{s},U)=0$, 
\begin{itemize}
\item in the case of \textit{bosons}, if\\
\begin{align}\label{eq:SuppLawBosons}
\prod_{\alpha=1}^N \Lambda_\alpha(\vec{s})\neq 1;
\end{align}

\item in the case of \textit{fermions}, if\\
\begin{align}\label{eq:SuppLawFermions}
\Lambda(\vec{s})\neq \Lambda_\mathrm{ini}.
\end{align}
\end{itemize}

For both particle types, the suppression of a final particle configuration $\vec{s}$ can be immediately determined from the corresponding final eigenvalue distribution. This entails drastically reduced computational costs compared to the conventional method via the evaluation of the permanent or determinant of $M$. Because the eigenvalues are integer roots of unity, only those bosonic final particle configurations are allowed for which the phases of the final eigenvalue distribution add up to zero. For fermions, on the other hand, the final eigenvalue distribution of non-vanishing configurations is uniquely defined by $\Lambda_\mathrm{ini}$. Note that the trivial permutation $\mathscr{P}=\unit$ only has eigenvalues $1$ so that conditions~\eqref{eq:SuppLawBosons} and \eqref{eq:SuppLawFermions} can never be fulfilled. Thus, unitaries of the form~\eqref{eq:Ueigen} must exhibit some non-trivial permutation characteristics~\eqref{eq:MatrixSym} in order for a suppression law to take effect.

\begin{figure}[t]
\centering
\includegraphics[width=0.47\textwidth]{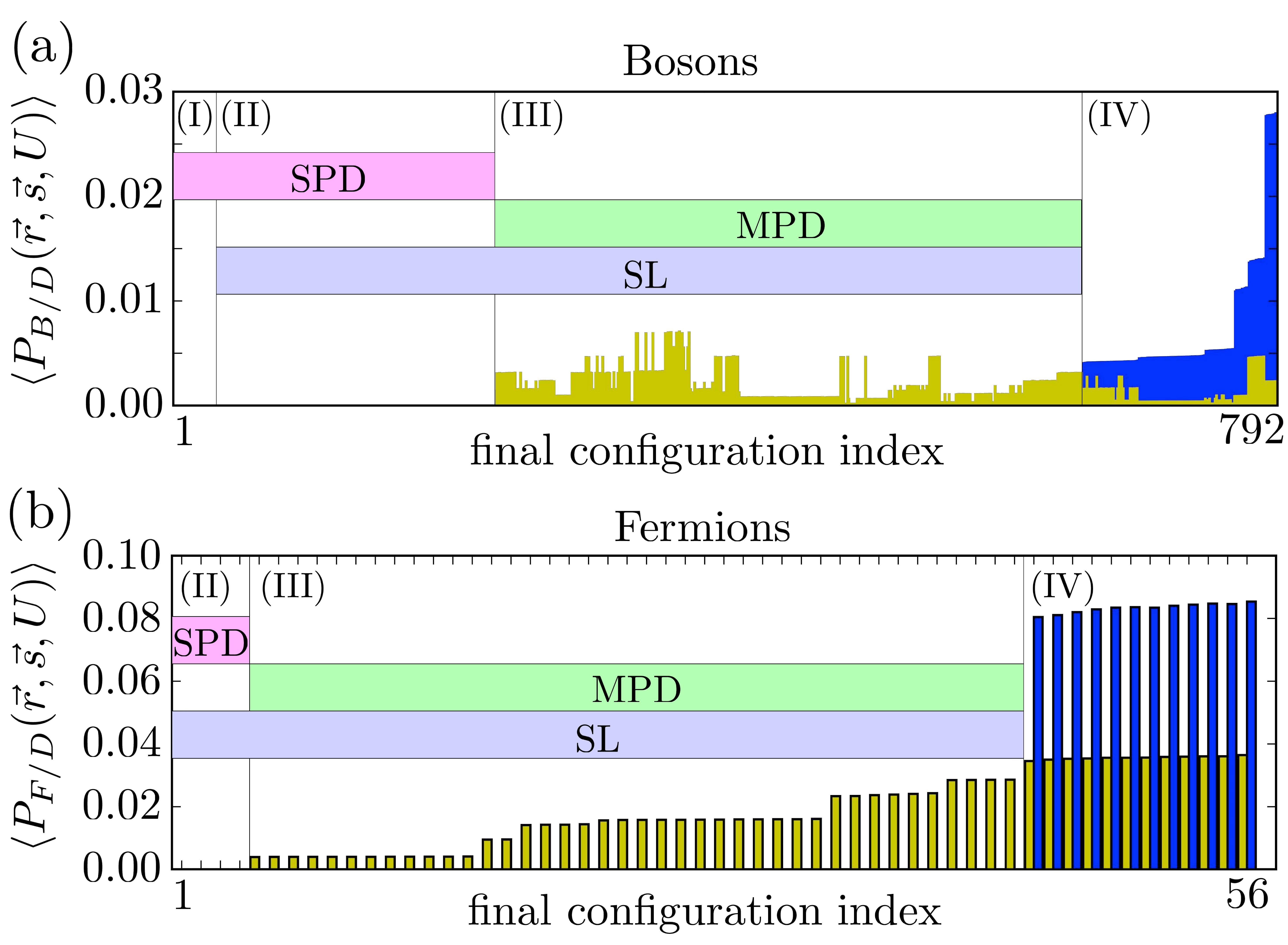} 
\caption{(Color online) Numerical evaluation of mean transmission probabilities $\langle P_\mathrm{B/F/D}(\vec{r},\vec{s},U)\rangle$ for the example illustrated in Fig.~\ref{fig:eigenvalues}, with bosonic (B) and fermionic (F) event probabilities depicted by blue bars in panels (a) and (b), respectively. Yellow bars represent event probabilities for distinguishable (D) particles, renormalised on the set of singly occupied modes in (b). Data were generated by averaging over $10\,000$ randomly chosen eigenbases of the permutation operator $\mathscr{P}$, for all output configurations of bosons, fermions and distinguishable particles, respectively. Forbidden output events are subdivided in event classes (I, II, III), where suppression is due to single-particle dynamics (SPD) in classes (I, II), and due to many-particle dynamics (MPD) in (III). \emph{All} suppressed events of classes (II, III) are predicted by our suppression laws (SL)~(\ref{eq:SuppLawBosons}, \ref{eq:SuppLawFermions}), while those SPD-suppressed events in (I) need some special consideration, see \cite{Dittel-TDarticle-2017}. From altogether $792$ bosonic and $56$ fermionic transmission events the subsets of those with nonvanishing detection probabilities are collected in set (IV), in increasing order.}
\label{fig:RandomSim}
\end{figure}

As an example highlighting the applicability of the above suppression laws, we further inspect the example of Fig.~\ref{fig:eigenvalues}: By rotations in the degenerate subspaces of $\mathscr{P}$ \footnote{Recall our discussion below Eq.~\eqref{eq:invariant}, and note that the columns of the generated eigenbases $A$ are assigned to the same eigenvalues for each realisation}, we numerically generate $10\,000$ random eigenbases $A$ and calculate the mean transition probability $\langle P_\mathrm{B/F/D}(\vec{r},\vec{s},U)\rangle$ for bosons, fermions and distinguishable particles. Figures~\ref{fig:RandomSim}~(a) and (b) illustrate the probability distributions for indistinguishable bosons and fermions, respectively, together with those for distinguishable particles. In both cases, there is a domain (III) that reveals a suppression of final particle configurations due to many-particle interference. All these suppressed final configurations, as well as those in domain (II) which are associated with single particle dynamics, are fully accounted for by our suppression laws. Note that configurations $\vec{s}$ in domains (I) and (II) occur with zero probability, irrespective of the particle type and mutual distinguishability. This results from zeros in $U$, that is from single-particle dynamics rather than many-particle interference. A detailed study of these configurations is provided in \cite{Dittel-TDarticle-2017}.

Let us now apply the above formalism to the particular unitary 
\begin{align}
U^\mathrm{FT}_{j,k}=\frac{1}{\sqrt{n}}\exp\left(i\frac{2\pi}{n}(j-1)(k-1)\right)
\end{align}
of the discrete Fourier transformation, which was thoroughly studied in the context of suppression laws \cite{Tichy-ZT-2010,Tichy-MP-2012,Carolan-UL-2015,Crespi-SL-2016,Su-MI-2017}. For any (cyclic) permutation $\pi^\mathrm{FT}(j)=1+\modu[j+\frac{n}{m}-1,n]$ of order $m$, such that $n/m\in\mathbb{N}$, we find a mode-exchange symmetry given by $U^\mathrm{FT}_{\pi(j),k}=U^\mathrm{FT}_{j,k}\exp\left(i\frac{2\pi}{m}(k-1)\right)$. By comparison with Eq.~\eqref{eq:MatrixSym}, we infer $Z=\unit$ and read off the eigenvalue $\lambda_k=\exp(i\frac{2\pi}{m}(k-1))$ associated with the $k$th eigenvector, i.e. the $k$th column of $U^\mathrm{FT}$. Application of the above suppression law for indistinguishable bosons then reveals that, for initial states invariant under $\pi^\mathrm{FT}$, all final particle configurations for which $\prod_{\alpha=1}^N\Lambda_\alpha(\vec{s})\neq 1$ must be suppressed. This coincides exactly with the findings of \cite{Tichy-MP-2012}. In the case of fermions, we simply have to consider the single cycle length $m$, such that $\Lambda_\mathrm{ini}$ contains every eigenvalue (i.e. each $m$th root of unity) exactly $N/m$ times. Accordingly, all final states vanish whose final eigenvalue distribution does not coincide with $\Lambda_\mathrm{ini}$. This condition provides a stronger criterion than the one given in \cite{Tichy-MP-2012}, where only those final particle configurations for which $\prod_{\alpha=1}^N \Lambda_\alpha(\vec{s})\neq (-1)^w$, with $w$ the number of transpositions necessary to permute $\vec{r}$ according to $\mathscr{P}$, are predicted to be suppressed. This is only a subset of the states suppressed according to our new suppression law~\eqref{eq:SuppLawFermions}, which, thus, expands the known fermionic suppression law for the discrete Fourier transformation. A detailed analysis of the Fourier suppression law in the present formalism is provided in \cite{Dittel-TDarticle-2017}, where we additionally elaborate upon \emph{all} other suppression laws so far described in literature.

In principle, the suppression criteria described above necessitate perfect indistinguishability of all particles, and unitaries obeying exactly the mode-exchange symmetry as specified by Eqs.~\eqref{eq:Ueigen} and \eqref{eq:MatrixSym}. In realistic experiments, however, deviations will arise. In principle, one can calculate the impact of such deviations  explicitly by unitary reconstruction \cite{Laing-SS-2012,Rahimi-DC-2013,Tillmann-UR-2016,Spagnolo-LU-2016} and an accurate treatment of partial distinguishability in the probability calculations \cite{Shchesnovich-SC-2014,Shchesnovich-PI-2015,Tichy-SP-2015,Tillmann-GM-2015,Khalid-PS-2017}. Clearly, making full use of this machinery will forfeit the computational advantage offered by our analytical suppression laws. Therefore, we here resort to a stability analysis with respect to small deviations from the ideal situation. Regarding imperfect transformation matrices, we follow the procedure in \cite{Tichy-SE-2014} and model unitaries by $\mathcal{U}_{j,k}=U_{j,k}(1+\Delta_{j,k})$ where $U$ corresponds to the ideal transformation matrix, and $\Delta_{j,k}\in\mathbb{C}$ describe random deviations with zero mean. For an output particle configuration $\vec{s}_{\mathrm{sp}}$, which is suppressed in the ideal situation, $P_{\mathrm{B/F}}(\vec{r},\vec{s}_{\mathrm{sp}},U)=0$, and small inaccuracies, $|\Delta_{j,k}|\ll 1$, a first order approximation yields a deviation from perfect suppression according to
\begin{align}\label{eq:deviation-U}
\delta P_{\mathrm{B/F}}(\vec{r},\vec{s}_{\mathrm{sp}},\mathcal{U})\approx N \langle|\Delta|\rangle^2\frac{\prod_j s_j!}{\prod_k r_k!} P_{\mathrm{D}}(\vec{r},\vec{s}_{\mathrm{sp}},U)
\end{align}
with $\langle|\Delta|\rangle$ denoting the average absolute value of $\Delta_{j,k}$. This implies that the transition probability is only affected in second order in $\langle|\Delta|\rangle$ and weighted by the corresponding probability of fully distinguishable particles (which measures the magnitudes of the pertinent elements of $U$). However, the sensitivity to distortions increases linearly with $N$, and for indistinguishable bosons bunched events are preferentially emerging by virtue of the factor $\prod_j s_j!$. 

The effect of partial distinguishability is investigated in the tensor permanent approach as developed in Refs.~\cite{Shchesnovich-SC-2014,Shchesnovich-PI-2015,Tichy-SP-2015,Tichy-EE-2017}. In this method, particles in the same initial mode are considered fully indistinguishable. Distinguishability between particles in different modes is encoded in the distinguishability matrix $\mathcal{S}_{j,k}=\bracket{\Phi_j}{\Phi_k}$ with $\ket{\Phi_j}$ denoting the internal state of a particle in mode $j$, which accounts for all potentially distinguishing degrees of freedom. For fermions, we consider at most one particle per mode since a multiple occupation would require total distinguishability. For indistinguishable particles, $\mathcal{S}_{j,k}=1$ for all $j,k$, while for completely distinguishable particles, $\mathcal{S}=\unit$. Hence, we investigate small partial distinguishabilities as encoded by the matrix $\mathcal{S}_{j,k}=(1-\epsilon_{j,k})e^{i\eta_{j,k}}$, where $\epsilon_{j,k},\eta_{j,k}\in\mathbb{R}$ account for random deviations. In order to keep mutual distinguishabilities small, we assume $0\leq \epsilon_{j,k}\ll 1$ while the $\eta_{j,k}$ have zero mean and are bounded by $-\pi/2\ll\eta_{j,k}\ll\pi/2$. Again we denote the average (absolute) value by $\langle\epsilon\rangle$ and perform a first order expansion in $\epsilon$, resulting in
\begin{align}\label{eq:deviation-dist}
\delta P_\mathrm{B/F}^\mathrm{part}(\vec{r},\vec{s}_{\mathrm{sp}},U)\approx N \langle\epsilon\rangle P_{\mathrm{D}}(\vec{r},\vec{s}_{\mathrm{sp}},U)
\end{align}
for partially distinguishable bosons and fermions. Here, a linear dependence on the distortion parameter $\langle\epsilon\rangle$ arises, and the deterioration of the suppression is again weighted by the transition probability of distinguishable particles and the particle number. 

Both above scenarios of weakly disordered unitary evolution and of imperfect indistinguishability exhibit only a gradual deviation from perfect suppression. In this sense, the suppression laws here derived are robust with respect to such error sources. However, there are cases where the validity of the suppression laws may persist even in the presence of pronounced mutual distinguishability of the particles. These cases are elaborated upon detail in \cite{Dittel-TDarticle-2017}.


In summary, we have shown that the permutation symmetry of a many-particle Fock state determines a class of unitary transformation matrices which lead to a robust suppression of final particle configurations due to the interference of all involved many-particle paths. Our present approach accommodates all known unitary scattering scenarios featuring totally destructive many-particle interference within one formal framework. Our findings are in this sense not only comprehensive, but they also highlight and identify the interrelation between symmetry and totally destructive interference. Their relevance to both  bosons and fermions entails the applicability to a wealth of physical systems such as optical interferometers \cite{Carolan-UL-2015}, where any unitary scattering matrix can be implemented with discrete optical elements \cite{Reck-ER-1994}, or atomic interferometers \cite{Kaufman-TH-2018}, which are likely to host controlled complex many-body interference in the near future. This will aid any application-based design of many-particle scattering scenarios. We are thus confident that our results pave the way for new applications in the manipulation of many-particle quantum states. 

G.W., R.K. and C.D. acknowledge support by the Austrian Science Fund (FWF projects I 2562, P 30459 and M 1849) and the Canadian Institute for Advanced Research (CIFAR, Quantum Information Science Program). C.D. is receiving a DOC fellowship from the Austrian Academy of Sciences. G.D. and A.B. acknowledge support by the EU Collaborative project QuProCS (Grant Agreement No. 641277). Furthermore, G.D. is thankful to the Alexander von Humboldt foundation and M.W. is grateful for financial support from European Union Grant QCUMbER (no. 665148).

\bibliographystyle{aipnum4-1} 

%

\end{document}